\documentclass{ws-procs9x6}

\usepackage{upgreek}

\newcommand{\be}{\begin{equation}}
\newcommand{\ee}{\end{equation}}

\newcommand{\msum}{\sum_{m=0}^\infty {}^{{}^\prime}}
\newcommand{\nsum}{\sum_{n=0}^\infty {}^{{}^\prime}}
\newcommand{\im}{\mathrm{Im}}
\newcommand{\re}{\mathrm{Re}}
\newcommand{\tr}{\mathrm{Tr}}

\newcommand{\dyad}[1]{\mbox{\textbf{\textsf{#1}}}}
\newcommand{\bd}[1]{\mathbf{#1}}
\newcommand{\trans}{\mathrm{T}}

\newcommand{\kb}{k_\mathrm{B}}
\newcommand{\tlr}{\tilde{r}}
\renewcommand{\th}{\tilde{h}}
\newcommand{\tj}{\tilde{j}}

\newcommand{\nodisplay}[1]{}

\begin{document}

\title{CASIMIR-POLDER POTENTIAL IN THERMAL NON-EQUILIBRIUM}

\author{S. {\AA}. ELLINGSEN}
\address{Department of Energy and Process Engineering, Norwegian
University of Science and Technology, N-7491 Trondheim, Norway}
\author{Y. SHERKUNOV}
\address{Department of Physics, University of Warwick, Coventry CV4
7AL, United Kingdom}
\author{S. Y. BUHMANN and S. SCHEEL}
\address{Quantum Optics and Laser Science, Blackett Laboratory,
Imperial College London, Prince Consort Road, London SW7 2AZ, United
Kingdom}

\begin{abstract}
Different non-equilibrium situations have recently been considered when
studying the thermal Casimir--Polder interaction with a body. We show
that the Keldysh Green function method provides a very general common
framework for such studies where non-equilibrium of either the atom or
the body with the environment can be accounted for. We apply the
results to the case of ground state polar molecules out of equilibrium
with their environment, observing several striking effects. We
consider thermal Casimir--Polder potentials in planar configurations,
and new results for a molecule in a cylindrical cavity are reported, 
showing similar characteristic behaviour as found in planar geometry. 
\end{abstract}


\section{Introduction}

Casimir--Polder (CP) or retarded Van der Waals forces \cite{casimir48}
is the name given to electromagnetic dispersion forces between
electrically neutral, but polarisable particles (atoms, molecules) and
macroscopic objects. In the present paper we discuss CP potentials on
particles when the atom--body--environment system is not in thermal
equilibrium. There exists a rich literature on CP forces in thermal
equilibrium (cf.~the citations in Ref.~\refcite{buhmann07}), and
non-equilibrium systems of an excited atom inside planar structures
have been studied for a long time\cite{barton70, jhe91}. The latter
works showed that an atom excited to an energy higher than thermal
energies will have a spatially oscillating component in both force and
heating rate.

Two different types of non-equilibrium have recently been studied.
The first study concerned atoms next to a substrate whose temperature
differs from the environment temperature\cite{antezza05, antezza08}
where the present radiation was assumed to be practically unable to
excite the atoms. It was found that the imbalance between environment
and substrate temperature may lead to strong force components whose
sign depends on which of the two temperatures is greater. This
prediction was subsequently confirmed in an experiment by the group of
Cornell\cite{obrecht07}. 

The second study considered particles in an equilibrium thermal
background of bodies and environment at uniform temperature, but
prepared in an arbitrary superposition of internal eigenstates, and
thus not necessarily at equilibrium with this background
\cite{buhmann08} (cf.\ the similar results reported in
Ref.~\refcite{gorza07}; for an extension to non-uniform temperature
environments, see Ref.~\refcite{buhmann09}). It was demonstrated that
even ground state particles are subject to an oscillating force
component in the presence of macroscopic bodies at nonzero
temperature. While utterly unobservable for atoms (which are
essentially in their ground state when thermalized at room
temperature), there could be some hope of observing and even using
this effect for molecules\cite{ellingsen09a, ellingsen09b}, which have
excited states of very low energy. 

A similar investigation was recently made\cite{sherkunov09} by
applying Keldysh theory to the system of two atoms prepared in an
arbitrary state, in the presence of an external (thermal)
electromagnetic field. In the first half of this article, we discuss
the correspondence of the Keldysh formalism with both of the above
non-equilibrium theories (Sec.~\ref{Keldysh}).  In the second half
(Sec.~\ref{Sec2}), we apply the results for a particle in a
uniform-temperature environment to planar and cylindrical cavities.


\section{Correspondence of thermal non-equilibrium theories}
\label{Keldysh}

The recent studies by Antezza, Pitaevskii and
Stringari\cite{antezza05} (APS) on the one hand and Buhmann
and Scheel\cite{buhmann08} (BS) on the other hand both deal with
non-equilibrium situations. However, the theory of APS is concerned
with effects due to a thermal disequilibrium between different parts
of the environment. The theory of BS describes an in some sense
opposite situation where the focus is on a disequilibrium between the
atom in a non-thermalized state and its environment which is itself in
thermal equilibrium with uniform temperature.

In this section we demonstrate how the two situations may be bridged
through a non-equilibrium theory of CP forces based on the Keldysh
Green function method \cite{sherkunov09} by showing that the known
results for the BS and APS configurations can both be obtained within
this formalism. We restrict our derivation to the interaction between
a ground-state particle and a non-magnetic medium embedded in an
external (thermal) electromagnetic field, neglecting the
non-equilibrium dynamics. A more general situation will be considered
elsewhere.

The CP potential of a polarisable ground-state particle at
position $\mathbf{r}$ next to a non-magnetic body embedded in an
electromagnetic field can be calculated with the help of the Keldysh
Green function technique. We calculate the density matrix of the atom
using the Keldysh method\cite{keldysh65}. Resticting our
considerations to the potential at initial time, we may neglect terms
associated with atomic non-equilibrium dynamics. The shift of the
atomc ground-state due to the interaction with the electromagnetic
field can then be found in terms of the photon Green functions (for
details, see Ref.~\refcite{sherkunov09}]:
\begin{align}
U(\mathbf{r}) =&-\frac{1}{2\pi}\int_0^\infty d\omega\,
\omega^2
\tr\im\left\{
\bm{\alpha}(\omega)\cdot
\left[\hbar\mu_0
\dyad{G}^{(1)}(\mathbf{r},\mathbf{r},\omega)
-\bm{\rho}(\mathbf{r},\mathbf{r},\omega)
\right]\right\},
\label{keldysh}
\end{align}
where $\bm{\alpha}(\omega)$ is the ground-state polarizability of the
atom and $\dyad{G}^{(1)}(\mathbf{r},\mathbf{r},\omega)$ is the
scattering part of the Green tensor of the electromagnetic field,
$\dyad{G}(\mathbf{r},\mathbf{r}',\omega)$, which in turn is the unique
solution to 
\begin{equation}
\label{12}
\left[\bm{\nabla}\times\bm{\nabla}\times
 \,-\,\frac{\omega^2}{c^2}\,\varepsilon(\mathbf{r},\omega)\right]
 \dyad{G}(\mathbf{r},\mathbf{r}',\omega)
 =\bm{\delta}(\mathbf{r}-\mathbf{r}')
\end{equation}
together with the boundary condition at infinity. Finally, 
$\bm{\rho}(\mathbf{r},t;\mathbf{r}',t')
=-i\langle\hat{\mathbf{A}}(\mathbf{r}',t')
\hat{\mathbf{A}}(\mathbf{r},t)\rangle^\trans$, where
$\hat{\mathbf{A}}$ is the vector potential of the external
electromagnetic field. Note that the Keldysh result offers a clear
distinction between the influence of the zero-point fluctuations of
the field, the first term in Eq.~(\ref{keldysh}) and the contributions
from external fields as contained in the second term.

Let us first compare the Keldysh method with the BS
calculations\cite{buhmann08}. If the environment containing the body
and the external field is at global thermal equilirium with a uniform
temperature $T_E$, the density matrix $\bm{\rho}$ can be calculated
with the help of the fluctuation-dissipation theorem: 
\begin{equation}
\bm{\rho}(\mathbf{r},\mathbf{r}',\omega)
=-2i\hbar\mu_0N(\omega,T_E)\im
\dyad{G}(\mathbf{r},\mathbf{r}',\omega),
\label{FDT}
\end{equation}
where $N(\omega,T_E)=[\exp(\hbar\omega/k_\mathrm{B}T_E)-1]^{-1}$
denotes the thermal photon numbers.
Substituting this into Eq.~(\ref{keldysh}) and discarding the
position-independent contribution associated with the bulk Green
tensor ($\dyad{G}\mapsto\dyad{G}^{(1)}$), we find:
\begin{align}
U(\mathbf{r})=& 
-\frac{\hbar\mu_0}{2\pi} \int_0^{\infty}d\omega\,\omega^2
 [2N(\omega,T_E)+1]\tr\im\left[\bm{\alpha}(\omega)\cdot
 \dyad{G}^{(1)}(\mathbf{r},\mathbf{r},\omega)\right]
\notag \\
&+\frac{\hbar\mu_0}{\pi}\int_0^{\infty}d\omega\,\omega^2 
 N(\omega, T_E)\tr\left[\im\bm{\alpha}(\omega)\cdot
 \re\dyad{G}^{(1)}(\mathbf{r},\mathbf{r},\omega)\right].
\label{res1}
\end{align}
The first term in Eq.~(\ref{res1}) can be cast into an
alternative form by writing $\im z=(z-z^\ast)/(2i)$, using the
identities $\dyad{G}^\ast(\omega)=\dyad{G}(-\omega)$,
$\bm{\alpha}^\ast(\omega)=\bm{\alpha}(-\omega)$ and making the
substitution $\omega\mapsto -\omega$. The emerging integral over the
entire real frequency axis can be completed to a closed contour by
adding a vanishing integral over an infinite semi-circle in the
upper half of the complex frequency plane. Evaluating the contour
integral via Cauchy's theorem, we are left with the contributions from
the poles $i\xi_m$ of $[2N(\omega,T_E)+1]$, viz.
\begin{align}\label{U1}
U(\bd{r}) =&\mu_0\kb T\msum
\xi_m^2\tr\left[\bm{\alpha}(i\xi_m)\cdot  
\dyad{G}^{(1)}(\bd{r},\bd{r},i\xi_m)\right]\notag\\
&+\frac{\hbar\mu_0}{\pi}\int_0^{\infty}d\omega\,\omega^2 
 N(\omega, T_E)\tr\left[\im\bm{\alpha}(\omega)\cdot
 \re\dyad{G}^{(1)}(\mathbf{r},\mathbf{r},\omega)\right],
\end{align}
with $\xi_m=2\pi m\kb T/\hbar$. The prime at the Matsubara sum
indicates that the $m=0$ term carries half-weight. For an isotropic
particle, the ground-state polarizability in the perturbative limit
can be given as
\begin{equation}
\label{alpha}
\bm{\alpha}(\omega)=\lim _{\epsilon\rightarrow 0} \frac{1}{3\hbar}
\sum_k\biggl[
\frac{|\mathbf{d}_{0k}|^2}{\omega+\omega_k+i\epsilon}
-\frac{|\mathbf{d}_{0k}|^2}{\omega-\omega_k+i\epsilon}\biggr]
\dyad{I}
\end{equation}
[$\omega_k=(E_k-E_0)/\hbar$ are transition frequencies;
$\mathbf{d}_{0k}$ are electric dipole matrix elements; $\dyad{I}$ is
the unit tensor]. Using  
$\lim_{\epsilon\to 0}1/(x+i\epsilon)=\mathcal{P}/x-i\pi\delta(x)$
(with $\mathcal{P}$ principal value), the thermal CP potential of an
isotropic ground-state atom is given by
\begin{align}\label{U}
U(\bd{r}) =&\mu_0\kb T\msum
\xi_m^2\alpha(i\xi_m)  
\tr\dyad{G}^{(1)}(\bd{r},\bd{r},i\xi_m)\notag\\
&+\frac{\mu_0}{3}\sum_k |{\bf d}_{0k}|^2
\,\omega_k^2 N(\omega_k, T_E)\tr
 \re\dyad{G}^{(1)}(\mathbf{r},\mathbf{r},\omega_k).
\end{align}
The corresponding force $\mathbf{F}(\bd{r})=-\bm{\nabla}U(\bd{r})$
agrees exactly with the BS result for the force at initial time on a
ground-state atom in the perturbative limit, cf. Eq.~(25) of
Ref.~\refcite{buhmann08}. Cf. this reference for details of the
derivation and the dynamical and nonperturbative generalisation of
this result for arbitrary (incoherent) initial-state preparation of
the atom. The first term of Eq.~(\ref{U}) is the non-resonant force,
while the second term is a resonant term due to absorption processes,
which are pure non-equilibrium effects. Note that such processes will
unavoidably heat the atom, so that our result is only valid at initial
time when the atom is still in its ground state. When the atom has
become thermalized, the resonant term vanishes and the expression
reduces to a pure Matsubara-type sum similar to the non-resonant
expression. 

In the APS case \cite{antezza05}, the body is assumed to be held at a
uniform temperature $T_S$ different from that of the environment,
$T_E$, such that the total system is in a stationary non-equilibrium.
The atom is assumed to be at zero temperature in the sense that the
thermal energies $k_\mathrm{B}T_S,k_\mathrm{B}T_E$ are much smaller
than the energies $\hbar\omega_k$ necessary to excite the atom.
This means that terms $N(\omega,T_E)\im\bm{\alpha}(\omega)$, which are
proportional to the occupation numbers of photons at the atomic
frequencies, cf.~Eq.~(\ref{alpha}), can be neglected. The density
matrix $\bm{\rho}$ obeys the kinetic-like equation emerging in the
Keldysh method\cite{lifshitzpitaevskiiX}:
\begin{align}
&\bm{\rho}(\mathbf{r}, t;\mathbf{r}' ,t')
=\bm{\rho}_0(\mathbf{r},t,\mathbf{r}',t')
+\frac{\mu_0}{\hbar}\int\!d^3r_1\int\!dt_1\int\!d^3r_2\int\!dt_2\,
 [\bm{\rho}_0(\mathbf{r},t;\mathbf{r}_1,t_1)\nonumber\\ 
&\qquad\cdot\bm{\Pi}_A(\mathbf{r}_1,t_1;\mathbf{r}_2,t_2)
 \cdot\dyad{G}_A(\mathbf{r}_2,t_2;\mathbf{r}',t')\nonumber\\
&\quad+\dyad{G}_0(\mathbf{r},t;\mathbf{r}_1,t_1)
 \cdot\bm{\Pi}_R(\mathbf{r}_1,t_1;\mathbf{r}_2,t_2)
 \cdot\bm{\rho}(\mathbf{r}_2,t_2;\mathbf{r}', t')\nonumber\\
&\quad-\;\dyad{G}_0(\mathbf{r},t;\mathbf{r}_1,t_1)
 \cdot\bm{\Pi}_{12}(\mathbf{r}_1,t_1;\mathbf{r}_2,t_2)
 \cdot\dyad{G}_A(\mathbf{r}_2,t_2;\mathbf{r}',t')/(\mu_0\hbar)]; 
\label{rho} \\
&\dyad{G}(\mathbf{r},t;\mathbf{r}' ,t')
=\dyad{G}_{0}(\mathbf{r},t;\mathbf{r}',t')
 +\frac{\mu_0}{\hbar}\int\!d^3r_1\int\!dt_1\int\!d^3r_2\int\!dt_2\,
 \dyad{G}_{0}(\mathbf{r},t;\mathbf{r}_1,t_1)\nonumber\\ 
&\quad\cdot\bm{\Pi}_R (\mathbf{r}_1,t_1;\mathbf{r}_2,t_2)
 \cdot\dyad{G}(\mathbf{r}_2,t_2;\mathbf{r}', t'). \label{GR}
\end{align}
Here, the index 0 corresponds to free fields;
$\dyad{G}_A=\dyad{G}^\dagger$ is the advanced Green tensor; and
$\bm{\Pi}$ are polarisation operators defined in terms of fluctuating currents
$\bm{\Pi}_{12}(\mathbf{r},t;\mathbf{r'},t')=%
-i\langle\hat{\mathbf{j}}(\mathbf{r'},t')%
 \hat{\mathbf{j}}(\mathbf{r},t)\rangle^\trans$, 
$\bm{\Pi}_{21}(\mathbf{r},t;\mathbf{r'},t')=%
-i\langle\hat{\mathbf{j}}(\mathbf{r},t)%
\hat{\mathbf{j}}(\mathbf{r'},t')\rangle$, and $\bm{\Pi}_{R}(\mathbf{r},t;\mathbf{r'},t')=[\bm{\Pi}_{12}(\mathbf{r},t;\mathbf{r'},t')-\bm{\Pi}_{21}(\mathbf{r},t;\mathbf{r'},t')]\Theta(t-t')$, $\bm{\Pi}_{A}=\bm{\Pi}_{R}^{\dagger}$, where $\Theta$ is the Heaviside function. If the
body was in equilibrium with the environment, the solution of
Eq.~(\ref{rho}) would be given by Eq.~(\ref{FDT}). To account for
effects due to the temperature difference between the body and the
environment, we now solve Eq.~(\ref{rho}) iteratively starting with
the equilibrium solution~(\ref{FDT}). The first approximation reads:
\begin{align}
&\bm{\rho}^{(1)}(\mathbf{r},\mathbf{r}',\omega)
 =-2i\hbar\mu_0N(T_E,\omega) 
\im\dyad{G}(\mathbf{r},\mathbf{r}',\omega)
-\mu_0^2\int d^3r_1d^3r_2\,
 \dyad{G}_0(\mathbf{r},\mathbf{r}_1,\omega)
\nonumber\\
&\cdot\left\{N(T_E,\omega)
\bm{\Pi}_{21}(\mathbf{r}_1,\mathbf{r}_2,\omega)
-[N(T_E,\omega)+1]\bm{\Pi}_{12}(\mathbf{r}_1,\mathbf{r}_2,
\omega)\right\}
\cdot\dyad{G}^*(\mathbf{r}_2,\mathbf{r}',\omega).
\label{rho1}
\end{align}
Here, we have used Eq.~(\ref{GR}), the free-field density matrix
$\bm{\rho}_0(\mathbf{r},\mathbf{r}',\omega)%
=-2i\hbar\mu_0N(\omega,T_E)\im%
\dyad{G}_0(\mathbf{r},\mathbf{r}',\omega)$ and the
relation\cite{lifshitzpitaevskiiX}
$\bm{\Pi}_R-\bm{\Pi}_A=\bm{\Pi}_{12}-\bm{\Pi}_{21}$ for the
polarization operators. Resubstituting Eq.~(\ref{rho1}) into the
r.h.s.\ of Eq.~(\ref{rho}), we obtain a result of the
form~(\ref{rho1}) with $\dyad{G}_0 \mapsto
\dyad{G}_0+\dyad{G}_0\!\cdot\!\bm{\Pi}_R\!\cdot\!\dyad{G}_0$.
Repeating the iterations many times and performing the sum via a Born
series
$\dyad{G}=\dyad{G}_0+\dyad{G}_0\!\cdot\!\bm{\Pi}_R\!\cdot\!\dyad{G}_0
+\dyad{G}_0\!\cdot\!\bm{\Pi}_R\!\cdot\!\dyad{G}_0\!\cdot\!%
\bm{\Pi}_R\!\cdot\!\dyad{G}_0+\ldots$, we arrive at
\begin{align}
&\bm{\rho}(\mathbf{r},\mathbf{r}',\omega)=
 -\hbar\mu_0 \left[2iN(T_E,\omega) 
\im\dyad{G}(\mathbf{r},\mathbf{r}',\omega)+
\int d^3r_1d^3r_2\,\dyad{G}(\mathbf{r},\mathbf{r}_1,\omega)\right.
\nonumber\\
&\left.\cdot \left\{N(T_E,\omega)
\bm{\Pi}_{21}(\mathbf{r}_1,\mathbf{r}_2,\omega)
-[N(T_E,\omega)+1]\bm{\Pi}_{12}(\mathbf{r}_1,\mathbf{r}_2,
\omega)\right\}
\cdot\dyad{G}^*(\mathbf{r}_2,\mathbf{r}',\omega)\right].
\label{rhof}
\end{align}
As stated, the body is assumed to be at local thermal equilibrium with
temperature $T_S$. This allows us to implement the
fluctuation--dissipation theorem to calculate the polarisation
operators:
%
\begin{subequations}\label{pfdt}
\begin{align}
\bm{\Pi}_{12}(\omega, \mathbf{r},\mathbf{r'})
=&-\frac{i\hbar\varepsilon_0}{2\pi}\,\omega^2N(\omega,T_S) 
\im\varepsilon(\mathbf{r},\omega)
\bm{\delta}(\mathbf{r}-\mathbf{r'});\\
\bm{\Pi}_{21}(\omega, \mathbf{r},\mathbf{r'})
=&-\frac{i\hbar\varepsilon_0}{2\pi}\,\omega^2\,[N(\omega, T_S)+1] 
\im\varepsilon(\mathbf{r},\omega)
\bm{\delta}(\mathbf{r}-\mathbf{r'}).
\end{align}
\end{subequations}
Inserting Eqs.~(\ref{pfdt}) and (\ref{rhof}) into Eq.~(\ref{keldysh}),
we find:
\begin{align}
&U(\mathbf{r})=-\frac{\hbar\mu_0}{2\pi} \int_0^{\infty}d\omega\,
\omega^2 
[2N(\omega,T_E)+1]\im\tr\left[\bm{\alpha}(\omega)\cdot 
\dyad{G}^{(1)}(\mathbf{r},\mathbf{r},\omega)\right]
\notag \\
&\quad+\re\,\frac{\hbar\mu_0}{(2\pi)^2}\int_0^{\infty} d\omega\,
\frac{\omega^4}{c^2} 
\int d^3s\,\im\varepsilon(\mathbf{s},\omega)
\left\{ N(\omega, T_E)[N(\omega, T_S)+1]
\right. \notag \\
&\qquad\left.-[N(\omega, T_E)+1] N(\omega, T_S)\right\} 
\tr\left[\bm{\alpha}(\omega)\cdot
\dyad{G}(\mathbf{r},\mathbf{s},\omega)\cdot
\dyad{G}^\ast(\mathbf{s},\mathbf{r},\omega)\right].
\label{res}
\end{align}
The first term of Eq.~(\ref{res}) describes the equilibrium CP force,
cf.~Eq.~(\ref{res1}) above, the second term corresponds to the case
when the medium is not in equilibrium with the field. If the medium is
a homogeneous body of permittivity $\varepsilon(\omega)$ occupying a
volume $V_S$, then the latter leads to just the APS result for the
non-equilibrium force $\mathbf{F}_\mathrm{neq}(\mathbf{r})%
=-\bm{\nabla}U_\mathrm{neq}(\mathbf{r})$  [see Eqs.~(7) and (9) of
Ref.~\refcite{antezza05}]:
\begin{align}
\mathbf{F}_\mathrm{neq}(\mathbf{r})=
&\frac{\hbar\mu_0}{2\pi^2}\int_0^{\infty} d\omega\,
\frac{\omega^4}{c^2} 
\int_{V_S}d^3s
\left[\frac{1}{e^{\hbar\omega/k_\mathrm{B}T_S}-1}
-\frac{1}{e^{\hbar\omega/k_\mathrm{B}T_E}-1}\right ] 
\notag \\
 &\times
\im\varepsilon(\omega)\left\{\bm{\nabla}'
\re\tr\left[\bm{\alpha}(\omega)\cdot
\dyad{G}(\mathbf{r},\mathbf{s},\omega)\cdot 
\dyad{G}^\ast(\mathbf{s},\mathbf{r}',\omega)\right]
\right\}_{\mathbf{r}'=\mathbf{r}}.
\end{align}


\section{Thermal CP potential on a particle in uniform temperature
environment}
\label{Sec2}

To illustrate the non-equilibrium effects, we apply the general theory
of the previous section to specific scenarios. We will concentrate on
a non-equilibrium between the particle and the electromagnetic field
and consider the potential~(\ref{U}) of a ground-state particle in
an environment of uniform temperature.


\subsection{Planar systems}

The trace of the Green tensor at a distance $z$ to the right of a
half-space with reflection coefficients $r_s, r_p$ for $s,p$
polarization is\cite{tomas95}
\be\label{G}
  \tr\dyad{G}(\bd{r}, \bd{r},\omega) = \frac{i}{4\pi}\int_0^\infty
 \frac{qdq}{\beta}\left[\frac{2\beta^2c^2}{\omega^2}r_p
 -\sum_{\sigma=s,p} r_\sigma\right] e^{2i\beta z}
\ee
with $\beta=\sqrt{\omega^2/c^2-q^2}$. The integral over transverse
momentum $q$ naturally separates into a propagating part $q<\omega/c$
and an evanescent part $q>\omega/c$. 

\begin{figure}
 \begin{center}
  \psfig{file=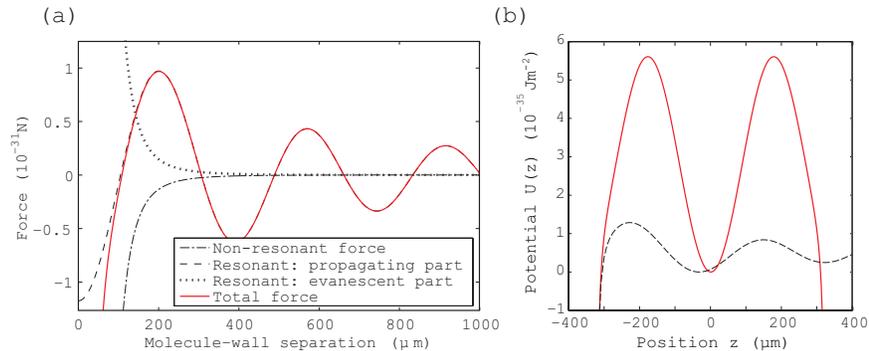, width = 4.5in}
  \end{center}
  \caption{(a) Components of the thermal CP force 
$F_z=-\partial U/(\partial z)$
on a ground state LiH molecule outside a gold half space at 300K.
(b) Enhanced potential from cavity of width $a=\lambda_{k0}$ (solid
line). The dashed line indicates the potential outside a single
half-space.}\label{fig:thermal}
\end{figure}

We use Eq.~(\ref{G}) to calculate the force on a ground state LiH
molecule outside a gold half-space at $T=300$K. The result is striking
(Fig.~\ref{fig:thermal}a): the evanescent part almost exactly
cancels the non-resonant part, and the propagating part is spatially
oscillating and dominates in the retarded regime.

Unfortunately, the spatially oscillating propagating force component
is very weak outside a half-space. We have investigated a scheme to
enhance the amplitude of the oscillating potential by fine-tuning the
width of a planar cavity to exactly one wavelength of light which
resonates with the dominating molecular transition\cite{ellingsen09b}.
Figure \ref{fig:thermal}b shows that the scheme works in principle, but
the enhancement factor thus achieved is not enough to bring the
oscillations into a regime which is likely to be observable. The
reason for this is primarily that the enhancement factor scales with
the logarithm of the Q-factor of the cavity (shown analytically and
numerically in Ref.~\refcite{ellingsen09b}) which strongly limits the
potentiality of such a scheme. 


\subsection{Cylindrical cavity}

Another candidate geometry is a molecule situated inside a cylindrical
cavity of radius $R$. At certain specific radii, resonances like that
for the planar cavity are expected to occur, where the two-dimensional
mode confinement holds the promise that resonant force components may
be much stronger than in the planar case. As a first step towards
exploiting this effect to realise strong thermal CP forces, we
investigate the spatial profile of resonant forces inside the
cylindrical cavity. Work on this problem is continuing. 

The trace of the Green tensor for points
$\bd{r}'=\bd{r}=(\rho,\theta,z)$ inside a cylindrical vacuum cavity
in an unbounded non-magnetic medium of permittivity $\varepsilon$
may, after much simplification, be written as\cite{li00,scheel08}
\begin{align}
  \tr \dyad{G}(\rho,\rho;\omega) =& \frac{i k}{2\pi}\int_0^\infty dt
\nsum \left\{(r_M + t^2 r_N)\right.\notag\\
  &\times\left.\left[\frac{n^2}{\phi^2 x^2}J_n^2(\phi x) 
+J_n^{\prime 2}(\phi x)\right]
+r_N \frac{x^2}{g^2}J_n^2(\phi x)\right\}.
\end{align}
Here, $x = g\sqrt{1-t^2}$, $g = kR$ and $k = \omega/c$. The
dimensionless radial co-ordinate is $\phi = \rho/R$ and the
integration variable $t$ is the dimensionless momentum component along
the cylinder axis, $h$, relative to $k$.\footnote[2]{For the
non-resonant term the dimensionless substitution variable $t=h/k$
cannot be used since $k=0$ for the zeroth Matsubara term. We use
$\tilde t=hR=gt$ in this case.} The reflection coefficients read
$r_{M,N} = - [H^{(1)}_n(x)/J_n(x)]\tlr_{M,N}$ with $\tlr_{M,N} =
(A+B_{M,N})/(A+B_{D})$ and
\begin{subequations}
\begin{align}
  A=& n^2[x^6-(2x_1^2+g^2)x^4+(2g^2+x_1^2)x_1^2x^2-g^2x_1^4],\\
  B_M=& g^2x_1^2x^2[\varepsilon \th_1^2x^2 
  - (\th_1\tj_2+\varepsilon\th_1\th_2)x_1x+\th_2\tj_2 x_1^2],\\
  B_N=& g^2x_1^2x^2[\varepsilon \th_1^2x^2 
  - (\varepsilon\th_1\tj_2+\th_1\th_2)x_1x+\th_2\tj_2 x_1^2],\\
  B_D=& g^2x_1^2x^2[\varepsilon \th_1^2x^2 
  - (\varepsilon+1)\th_1\tj_2x_1x+\tj_2^2x_1^2]
\end{align}
\end{subequations}
[$x_1= g\sqrt{\varepsilon -t^2}$; $x_2 = x$; 
$\tj_j = J'_n(x_j)/J_n(x_j)$; 
$\th_j = H^{(1)\prime}_n(x_j)/H^{(1)}_n(x_j)$].

In the perfectly conducting limit $|\varepsilon|\to \infty$ (at
nonzero $\omega$) we find
\be
  r_M\to -\frac{H_n^{(1)\prime}(x)}{J'_n(x)}\,;  
 ~~~~ r_N\to -\frac{H_n^{(1)}(x)}{J_n(x)}\,.
\ee
For the perfectly conducting cylinder, thus, the resonant radii for
radiation of a given frequency are $R_{ni}(\omega)$ and
$R'_{ni}(\omega)$ with $R_{ni}^{(\prime)}(\omega) = c
j_{ni}^{(\prime)}/\omega$ wherein $j_{ni}$ and $j_{ni}'$ are the $i$th
zero of $J_n(x)$ and $J_n'(x)$, respectively. When
$\varepsilon<\infty$ the resonances move away from these values.
Further details and analysis will be reported elsewhere.

\begin{figure}
 \begin{center}
  \psfig{file=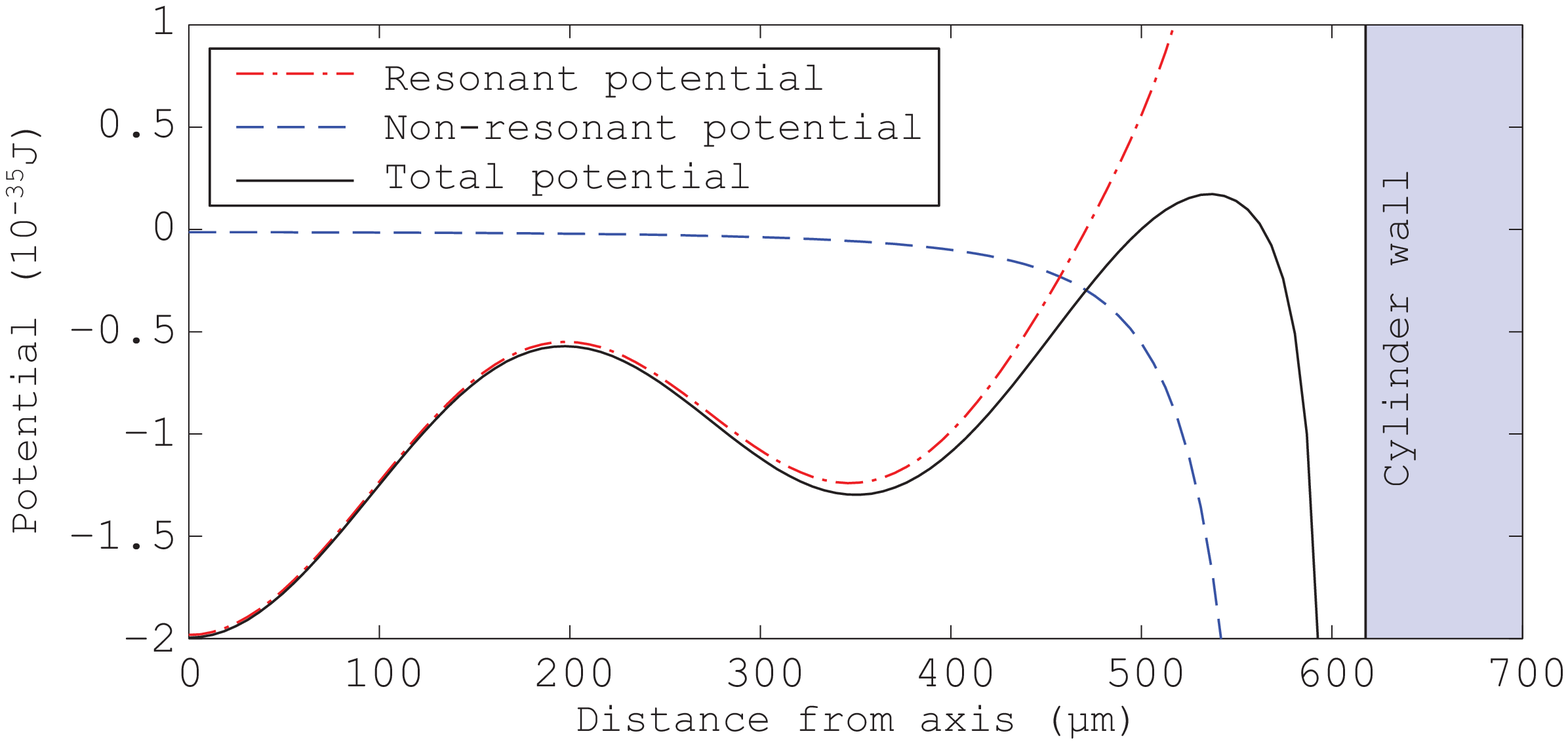, width = 3.5in}
  \end{center}
  \caption{The thermal CP potential on a LiH molecule in 
an infinitely thick cylindrical gold cavity at $T=300$K. The radius is
$R=1.5R_{11}\approx 618\upmu$m.}\label{fig:cyl}
\end{figure}

We plot the potential for the example of $R=1.5 R_{11}$ for
ground-state LiH in Fig.~\ref{fig:cyl}. For technical reasons, in the
cylindrical geometry splitting the resonant potential into propagating
and evanescent parts is no longer natural and straightforward. 
As a benchmark we have checked that the potential outside a plane is
regained close to the cylinder boundary in the limit of large cylinder
radius $R\gg 2\pi/k_{10}$.

Figure \ref{fig:cyl} shows clearly that the peculiar traits observed
for the resonant potential outside a half-space, depicted in
Fig.~\ref{fig:thermal}a, are present also in the cylindrical cavity as
one would expect. The resonant potential once again almost cancels the
non-resonant term close to the surface giving a resulting attractive
force in the near zone which is dramatically reduced compared to the
non-resonant term alone. As before the retarded regime is dominated by
oscillating behaviour. However, we also observe qualitative
differences from the planar case, e.g., the potential minimum closest
to the wall is no longer the deepest one as is the case in a planar
cavity\cite{ellingsen09b}. 


\section{Summary}

We have demonstrated that the recent complementary theories for the
thermal CP force between a ground-state particle and a body at
non-equilibrium between either the particle or the body and the
electromagnetic field may both be obtained using the Keldysh
formalism. Applying the results to planar and cylindrical geometries
of uniform temperature, we have found that even a ground-state
particle is subject to resonant force components which in both
geometries strongly cancel the well-known resonant force in the
nonretarded regime and lead to spatially oscillating forces for
retarded distances. The latter are expected to be much enhanced for
the cylindrical cavity, opening the perspective towards molecule
guiding with thermal photons.

\section*{Acknowledgments}

This work was financially supported partially by the UK Engineering
and Physical Sciences Research Council. We acknowledge financial
support by the European Science Foundation (ESF) within the activity
`New Trends and Applications of the Casimir Effect'
(\texttt{www.casimir-network.com}) and useful discussions with Dr.\
Mauro Antezza and Dr.\ Carsten Henkel. S.Y.B. is grateful for support
by the Alexander~von~Humboldt foundation. We finally thank Professor
Kimball Milton and the local organisers for an excellent conference.

\end{document}